\documentclass[prb,twocolumn,showpacs,superscriptaddress,amsmath,amssymb]{revtex4}
\usepackage{graphicx,color}% Include figure files
\usepackage{dcolumn}% Align table columns on decimal point
\usepackage{bm}% bold math

\begin{document}
%\preprint{APS/123-QED}
\title{Properties of atomic intercalated carbon $K_{4}$ crystals}

\author{Masahiro Itoh}
%\email{itoh.japan@gmail.com}
\affiliation{Institute of Multidisciplinary Research for Advanced Materials, Tohoku University, Aoba-ku, Sendai 980-8577, Japan}
\affiliation{Advanced Institute for Materials Research, WPI, Tohoku University, Aoba-ku, Sendai 980-8577, Japan}

%\author{Motoko Kotani}
%%\email{kotani@math.tohoku.ac.jp}
%\affiliation{Mathematical Institute, Graduate School of Science, Tohoku University, Aoba-ku, Sendai 980-8578, Japan}
%
\author{Seiichi Takami}
%\email{stakami@tagen.tohoku.ac.jp}
\affiliation{Institute of Multidisciplinary Research for Advanced Materials, Tohoku University, Aoba-ku, Sendai 980-8577, Japan}
%
%\author{Toshikazu Sunada}
%%\email{sunada@isc.meiji.ac.jp}
%\affiliation{Department of Mathematics, Meiji University, Tama-ku, Kawasaki 214-8571, Japan}
%
\author{Yoshiyuki Kawazoe}
%\email{kawazoe@imr.edu}
\affiliation{Institute for Materials Research, Tohoku University, Aoba-ku, Sendai 980-8577, Japan}
\author{Tadafumi Adschiri}
%\email{ajiri@tagen.tohoku.ac.jp}
\affiliation{Institute of Multidisciplinary Research for Advanced Materials, Tohoku University, Aoba-ku, Sendai 980-8577, Japan}
\affiliation{Advanced Institute for Materials Research, WPI, Tohoku University, Aoba-ku, Sendai 980-8577, Japan}
\date{\today}

\begin{abstract}
The stability of atomic intercalated carbon $K_{4}$ crystals, XC$_{2}$
(X=H, Li, Be, B, C, N, O, F, Na, Mg, Al, Si, P, S, Cl, K, Ca, Ga, Ge, As, Se, Br, Rb or Sr)
is evaluated by geometry optimization and frozen phonon analysis based on first principles calculations.
Although C $K_{4}$ is unstable, NaC$_{2}$ and MgC$_{2}$ are found to be stable.
It is shown that NaC$_{2}$ and MgC$_{2}$ are metallic and semi conducting, respectively.
\end{abstract}

\maketitle

\section{Introduction}

Recently, mathematical analysis elucidated that 120$^{\circ}$ bonding
forms a 3-D 10-member ring periodic structure ($K_{4}$ structure)
and can be regarded as the twin of the diamond structure\cite{Sunada}.
From the view point of carbon chemistry, 120$^{\circ}$ bonding is of the $sp^{2}$ bond type,
which form a 2-D structure with a 6-member ring, and the $K_{4}$ carbon structure has not been experimentally discovered.
Because of the wide applications of carbon materials (graphite, amorphous, diamond, CNT and fullerene\cite{Kawazoe}),
if such a new member exists in nature or can be synthesized, then this new material should have a very significant impact.

The stability and properties of this carbon $K_{4}$ structure
have been studied using first principles calculations\cite{RC-PRB-2008, IKNSKA-PRL-2009}.
It is interesting to note that carbon $K_{4}$ is metallic\cite{RC-PRB-2008, IKNSKA-PRL-2009},
although the phonon calculation results indicate the structure is unstable\cite{Yao}.
Compared with the graphite structure of $sp^{2}$ bonding, the binding energy per bond is slightly lower.
This is possibly one of the reasons for the instability of the carbon $K_{4}$ crystals against the thermal vibration.

It is well understood that atomic intercalation to the carbonaceous materials
changes the structural stability and sometimes provides totally different properties.
We propose that atomic intercalation will affect the electronic distribution in the $K_{4}$ crystals and possibly stabilize the structure.
Over 40 years ago, disilicides such as SrSi$_{2}$ were found to have $K_{4}$ type flame of Si\cite{SrSi2}.
This result encouraged us to investigate impurity intercalated carbon $K_{4}$ systems
although such crystal structures are not found in the carbide system\cite{carbides-review}.

This study examines the structural stability and several properties of the atomic intercalated carbon $K_{4}$ crystals,
XC$_{2}$ (X=H, Li, Be, B, C, N, O, F, Na, Mg, Al, Si, P, S, Cl, K, Ca, Ga, Ge, As, Se, Br, Rb or Sr) by first principles calculations.

\section{Computational Methods}

First principles calculations based on density functional theory\cite{HK, KS}
are performed for the $K_4$ type crystals XC$_{2}$, C, Si and SrSi$_{2}$
using the Vienna Ab-initio Simulation Package (VASP)\cite{Kresse}.

Local density approximation (LDA)\cite{PZ, CA} is used for the exchange-correlation energy functional.
In the present calculations, all these crystals are considered as spin-un-polarized systems.

To reduce the computational costs, the projector-augmented wave method\cite{PAW} is used to approximate
electrons in each atom in the crystal.
For H, Li, Be, B, C, N, O, F, Na, Mg, Al, Si, P, S, Cl, K, Ca, Ga, Ge, As, Se, Br, Rb and Sr,
valence electrons of 1$s^{1}$,
2$s^{1}$, 2$s^{2}$, 2$s^{2}$2$p^{1}$, 2$s^{2}$2$p^{2}$, 2$s^{2}$2$p^{3}$,
2$s^{2}$2$p^{4}$, 2$s^{2}$2$p^{5}$,
3$s^{1}$, 3$s^{2}$, 3$s^{2}$3$p^{1}$, 3$s^{2}$3$p^{2}$, 3$s^{2}$3$p^{3}$,
3$s^{2}$3$p^{4}$, 3$s^{2}$3$p^{5}$,
3$s^{2}$3$p^{6}$4$s^{1}$, 3$s^{2}$3$p^{6}$4$s^{2}$, 4$s^{2}$4$p^{1}$, 4$s^{2}$4$p^{2}$, 4$s^{2}$4$p^{3}$,
4$s^{2}$4$p^{4}$, 4$s^{2}$4$p^{5}$,
4$s^{2}$4$p^{6}$5$s^{1}$ and 4$s^{2}$4$p^{6}$5$s^{2}$ are considered, respectively.

To evaluate the stability of the  XC$_{2}$ $K_{4}$ crystal structure, the following procedure is adopted.

(1) The original conventional unit cell for the carbon $K_{4}$ crystal is shown in Figure 1(a).
Four X atoms (X = H, Li, Be, B, C, N, O, F, Na, Mg, Al, Si, P, S, Cl, K, Ca, Ga, Ge, As, Se, Br, Rb, or Sr) are allocated
to reduced coordinates ($\frac{1}{8}$, $\frac{1}{8}$, $\frac{1}{8}$), ($\frac{7}{8}$, $\frac{5}{8}$, $\frac{3}{8}$),
($\frac{3}{8}$, $\frac{7}{8}$, $\frac{5}{8}$), ($\frac{5}{8}$, $\frac{3}{8}$, $\frac{7}{8}$) in the original unit cell.
The primitive unit cell of XC$_{2}$ with $P4_{3}32$ ($O^{6}$) symmetry is shown in Fig. 1(b).
The symmetric plane of the atomic intercalated system is shown in Fig. 1(c).
As shown in Fig. 1(d), three adjacent atoms (red) are aligned on a straight line with the same distance (C-X, and X-C).

(2) Binding energy versus volume curve is evaluated and fitted using Murnaghan's equation of state\cite{Murnaghan}.
For these calculations,
Brillouin zone integration is performed for 8$\times$8$\times$8 \textit{\textbf{k}}-point meshes generated
by the Monkhorst-Pack scheme.
The residual minimization/direct inversion in the iterative subspace method
is used to accelerate the convergence of self-consistent total energy calculations.
The convergence criterion is set to be within 1 $\times 10^{-8}$ eV/formula unit cell.
The cut-off energy for the plane-wave expansion of valence electrons for the primitive unit cell is determined,
so that the number of plane waves is constant over a full range of the lattice constant.
Around the minimum of binding energy versus volume curve,
the cut-off energies were set to be 500 and 400 eV for C and XC$_{2}$, and Si and XSi$_{2}$ crystals, respectively.

(3) The unit cells obtained from step (2) are optimized with freedom of the atomic configuration under the symmetry constraint.

(4) The unit cells obtained from step (3) are optimized without any constraint on the crystal structure.
The convergence criterion is set to be within 1 $\times 10^{-7}$ eV/\AA\ unit cell.

(5) The frozen phonon calculations were performed using the FROPHO code\cite{FROPHO}
which is based on Parlinski-Li-Kawazoe method\cite{Parlinski-Li-Kawazoe}.
To obtain the force constants for the phonon calculations, the atomic displacements are set to be 0.01 \AA\ .
The Born - von Karman boundary condition is applied for each primitive unit cell obtained with multiplication
after steps (3) and (4) for the phonon calculation.
In these calculations, the primitive cells are multiplied by 2 for each direction of the unit vector.

\section{Results and Discussions}

The structure with $K_{4}$ crystal flame considered for C, XC$_{2}$, Si and SrSi$_{2}$ systems in the present study
maintain $K_{4}$ type crystal flame composed of C or Si with distortion through the whole process of the geometry optimization.
Furthermore, the crystal symmetry is maintained within an accuracy of less than 0.001\AA\ .
The lattice constant ($a$), volume at lattice constant ($V_{0}$), cohesive energy per atom ($E_{coh}$) 
and bulk modulus at $V_{0}$ ($B_{0}$) are evaluated for the fully optimized structure.
In this study, $E_{coh}$ for XC$_{2}$: $E_{coh, XC_{2}}$ is defined as
\begin{eqnarray}
E_{coh, XC_{2}}
\nonumber &\equiv\frac{-E_{XC_{2}}+N_{X atom}\times E_{X atom}+N_{C atom/K_{4}}\times E_{C atom}}{{N_{X atom}+N_{C atom/K_{4}}}}.
\end{eqnarray}
Here, $E_{A}$ and $N_{A}$ are the total energy and the number of atoms for A, respectively.
To obtain these values, Murnaghan's equation of state is used to fit the binding energy versus volume curve.
Parameters in the equation are determined using the least squares method in the range of about $0.8V_{0} < V < 1.2V_{0}$,
where the root mean square is set to be less than 5.0 meV/atom.

The obtained values of $a$, $V_{0}$, $E_{coh}$, and $B_{0}$ for the crystals are shown in Figure 2(a).
In XC$_{2}$, the values of $a$ and $V_{0}$ generally increase with the atomic number of X.
Generally, the values of $a$ for the XC$_{2}$ crystals are larger than that for carbon $K_{4}$.
That is, generally the carbon $K_{4}$ flames are expanded by intercalating X to the carbon systems.

However, the values of $V_{0}$ for X=H, Li, ..., S, are smaller than that of carbon $K_{4}$ while those for X=Cl, ..., Sr are larger.
This is because of the sufficiently wide vacant spaces in carbon $K_{4}$ that is compensated by X atoms
and the increase of the occupied spaces by X atoms for the increase of the atomic number of X.

On the other hand, the values of $E_{coh}$ decrease with the increasing period of X.
There is a strong negative correlation with $a$ and $V_{0}$ as shown
in the correlation coefficients: $r_{a-E_{coh}}$=-0.792 and $r_{V_{0}-E_{coh}}$=-0.796.
For each period of X in the XC$_{2}$, $E_{coh}$ generally shows largest values for intermediate elements X: IIIB or IVB.
This enhanced energy gain can be explained by the completion of the electronic shell in C atoms composing $K_{4}$ type flame,
the bonds formation with X in the XC$_{2}$ crystals and the bonds expantion which weaken the bonds. 
The $E_{coh}$ values of the XC$_{2}$ are not larger than that of the carbon $K_{4}$.
This is apparently resulted from the relative weakness of the X-C bonds in the XC$_{2}$ crystals
compared with the C-C bonds in the carbon $K_{4}$ crystal.
Subsequently, the absence of correlation between $E_{coh}$ and dynamical stability is shown.

The values of $B_{0}$ also decrease with the increasing period of X, similar to the case of $E_{coh}$.
The $B_{0}$s of XC$_2$ are generally smaller than that of C.
There is a strong correlation between $B_{0}$ and $E_{coh}$ as indicated by the correlation coefficient: $r _ { E_{coh} - B_{0} }$=0.848.

From the values of $a$ and $V_{0}$ in Si and SrSi$_{2}$,
the relation shown in the C and XC$_{2}$ systems also seems to hold for the Si and XSi$_{2}$ systems.
However, the values of $E_{coh}$ and $B_{0}$ of SrSi$_{2}$ are larger than those of Si.
This relation is apparently different from the cases of C and XC$_{2}$ systems.
This can be attributed to the X-Si bond strength being much higher than that of the Si-Si bonds in XSi$_{2}$.

To understand the correlation between the stability and the structure in detail,
the nearest-neighbour distances between different atoms specified in Fig. 1(e) 
$d$ C-C,  $d$ X-C$_A$, $d$ X-C$_B$, $d$ X-C$_C$, $\overline{d}$ X-C (the average of all $d$ X-C) and $d$ X-X
as well as the angle $\angle$ X-C$_{A}$-C$_{B}$,
and the minimal dihedral angle for the nearest-neighbour C atoms $\angle$ C-C-C-C are investigated.

As shown in Fig. 2(b), in general $d$ C-C, $\overline{d}$ X-C and $d$ X-X increase with the increasing atomic number in X,
similar to the cases of $a$ and $V_{0}$.
The $d$ C-C and $\overline{d}$ X-C show very strong correlation with the $a$,
indicated by the correlation coefficients: $r_{a-d_{C-C}}$=0.930 and $r_{a-\overline{d}_{X-C}}$=0.987.
These distances have also correlation with $E_{coh}$:
$r_{E_{coh}-d_{C-C}}$=-0.630 and $r_{E_{coh}-\overline{d}_{X-C}}$=-0.741, although the values are smaller.

It is apparent that the correlations shown in $\overline{d}$ X-C with $a$ and $E_{coh}$ are
resulted from the very strong correlation of $d$ X-C$_{A}$ and $d$ X-C$_{B}$
with $a$ and $E_{coh}$ as indicated by the correlation coefficients:
$r_{a-d_{X-C_{A}}}$=0.824, $r_{a-d_{X-C_{B}}}$=0.973, $r_{a-d_{X-C_{C}}}$=0.180,
$r_{E_{coh}-d_{X-C_{A}}}$=-0.759, $r_{E_{coh}-d_{X-C_{B}}}$=-0.796, and $r_{E_{coh}-d_{X-C_{C}}}$=-0.051.

Here, it should be noted that the each inequivalent atomic distance between X and C:
$d$ X-C$_{A}$, $d$ X-C$_{B}$, and $d$ X-C$_{C}$ show interesting X dependence.
The XC$_{2}$ systems with X of 3rd, 4(5)th periods show distinct differences between X=IA and IIA, and other Xs.
In the compound with X=IA or IIA, the $d$ X-C$_{A}$ and $d$ X-C$_{B}$ are relatively larger
while $d$ X-C$_{C}$ is smaller compared with those of the other compounds with X of the neighbour atomic number.
The angle $\angle$ X-C$_{A}$-C$_{B}$ show distinct smaller values for the compounds with X=IA and IIA
and this is attributed to the X dependence of the $d$ X-C$_{A}$ and $d$ X-C$_{B}$.
The dihedral angle $\angle$ C-C-C-C shown in Tab. II also show smaller values for the compounds with X=IA or IIA
although the differences with other compounds are very small.
Apparently, those differences are due to the number of the outer most valence electrons in X.
However, an exceptional trend is shown in the systems with X of 2nd period for those values
and the trend can be attributed to the elemental closeness between the X and C.

The relative position of X in the crystal is maintained resulting in $r_{a-d_{X-X}}$=1.000.
Therefore, $d$ X-X can be considered as a standard distance in the XC$_{2}$ systems.
To emphasize the similarity of the structures, various distance ratios of the nearest-neighbour distances to $d$ X-X are shown in Fig. 2(c).
Although significant larger (smaller) $d$ C-C / $d$ X-X, $\overline{d}$ X-C / $d$ X-X, and $d$ C-C$_{C}$ / $d$ X-X ($d$ C-C$_{A}$ / $d$ X-X)
are shown in the crystals with X=C, N, and O, these ratios are almost constant in the same group of X.

In Figure 3, the phonon density of states(DOS) without step (4) for C and XC$_{2}$
(X=H, Li, Be, B, C, N, O, F, Na, Mg, Al, Si, P, S, Cl, K, Ca, Ga, Ge, As, Se, Br, Rb or Sr), Si and SrSi$_{2}$
with $K_{4}$ type flame of C or Si are depicted.
From the above discussions and Fig. 2(a), the intermediate value of $a$ realized for the compounds with X in 3rd (4th) (4.5-4.8 \AA\ )
and the accompanied interatomic distances in these crystals seem to make the band narrower.
Namely, the inter atomic distances roughly determine the vibrational frequency range in these crystals.
The imaginary frequency modes rarely appear in NaC$_{2}$, MgC$_{2}$ and SrSi$_{2}$
although they appear in relatively wide imaginary frequency range in C and Si.

For the structures of C, NaC$_{2}$, MgC$_{2}$, Si and SrSi$_{2}$ obtained after step (4),
the phonon DOS and dispersion relationship are also evaluated.
As shown in Fig. 4, the phonon DOS for the fully optimized structures of the other XC$_{2}$ crystals show
almost the same shapes as those of the optimized structures from step (3).
Therefore, it is expected that the discussion in the previous paragraph is significant
and skipping step (4) is expected to be effective for reducing the computational cost in this study.

As shown in Figure 4, in carbon $K_{4}$, although the imaginary frequency mode is not appear at the $\Gamma$ point,
it appear in wide range of other $\textbf{\textit{k}}$ points as indicated by Yao \textit{et al} \cite{Yao}.
On the other hand, silicon $K_{4}$ has imaginary modes even at the $\Gamma$ point.
Namely, the carbon and silicon $K_{4}$ crystal structures are on a high rank of saddle points in those potential energy surfaces
and the lifes of them are expected to be very short in nature.

On the other hand, in our calculations although NaC$_{2}$, MgC$_{2}$ and SrSi$_{2}$ have imaginary frequencies
in the acoustic modes slightly around the $\Gamma$ point, there are no imaginary modes in the other $\textbf{\textit{k}}$ points.
These suggest that those structures are stable although they break the structures against the phonon vibration with long wave length limit.
The existence of the SrSi$_{2}$ $K_{4}$ in nature\cite{SrSi2} suggests that the results
in this study based on the density functional theory can predict the thermal stability of the structure.
From this fact and the calculation results, the stable existence of $K_{4}$ type crystal structures in NaC$_{2}$ and MgC$_{2}$ is expected in nature
although they do not realize maximum $E_{coh}$ in the considered XC$_{2}$ crystals.

From the maximum frequencies obtained for diamond (1300 cm$^{-1}$ for C\cite{Tohei}, 500 cm$^{-1}$ for Si\cite{Giannozzi})
and graphite (1600 cm$^{-1}$ for C)\cite{Tohei},
the orders of both C-C and Si-Si chemical bonds can be considered less than 1, as discussed by Yao \textit{et al}\cite{Yao}.
Therefore, based on the rough expectation from the mono atomic systems,
the orders are expected to be less than 1 between the atoms composing $K_{4}$ type flame of NaC$_{2}$, MgC$_{2}$ and SrSi$_{2}$ crystals.
The characters of chemical bonds are analysed from the valence charge distribution as described below.

To understand the stability of the $K_{4}$ systems, the valence charge distributions are analysed.
Fig. 5 shows various types of distribution in the conventional or the primitive unit cells
for the fully optimized C, NaC$_2$, MgC$_2$, Si and SrSi$_2$ crystals with $K_4$ type flame.
Fig. 5(a) details planes selected to show the distribution of quantities noted in Figs. 5(c), (e), and (g).
The populations around the Sr atoms are neglected for easier comparison in Figs. 5(c) and (e).

Figures 5(b) and (c) show the isosurface and contour of charge density.
Apparently, the valence charge is locally distributed along the lines
between adjacent C or Si atoms of the $K_{4}$ type flame for the considered systems.
It should be noted that a distinctive difference appears between the charge distribution
of the C(XC$_{2}$) and Si(SrSi$_{2}$) crystals in the intermediate region
between the nearest-neighbour atoms in those flames. The charge is more localized to the C than Si atoms.
These features are also indicated in the diamond crystal structures as shown in the previous studies\cite{Cohen}.
It can be said that the atomic intercalation change the distribution slightly.
However, as discussed for the phonon frequencies, this quantity is sufficient to change the properties of the crystals.

Figures 5(d) and (e) shows the differences between the valence charge density of the crystals (ex. $\rho_{NaC_{2}}(\textbf{r})$)
and their separated components: the intercalated atoms (ex. $\rho_{Na}(\textbf{r})$) and the flame (ex. $\rho_{C_{2}}(\textbf{r})$).
As shown in Fig. 5(d), the significant excess charge is shown in the region between the adjacent Na-C, Mg-C and Sr-Si
for NaC$_{2}$, MgC$_{2}$ and SrSi$_{2}$, respectively; therefore suggesting bond creation in those regions.
Furthermore, as shown in Fig. 5(e), in NaC$_2$, MgC$_2$ and SrSi$_2$,
the charge accumulations increase in the region of $\pi$ type orbitals while those decrease along the $\sigma$ type bonds;
suggesting that the $\pi$ type bonds are strengthened and the $\sigma$ type bonds are weakened.
On the other hand, the maximum frequency of the phonon modes shown in Fig. 4 is less than the mono atomic systems.
Therefore, the net bond orders in the $K_{4}$ type flame are expected to be decreased by the atomic intercalation.

Here, the nature of the bonds is noted from the information of the ELF.
Figs. 5(f) and (g) show the isosurface (0.8 [-]) and the contour.
As shown here, C, NaC$_2$, MgC$_2$, Si and SrSi$_{2}$ has similar ELF distribution between the adjacent C-C and Si-Si.
However, the feature of $\pi$ type bonds formation is strengthened with significance.
The relatively smaller value of $\angle$ C-C-C-C ($\angle$ Si-Si-Si-Si) seems to have correlation with this feature.
The strengthened $\pi$ type bonds seem to stabilize the structures against the themal vibration.
From the comparison with the case of graphite, the absense of the $\pi$ type bonds indicated
by Yao \textit{et al.}\cite{Yao}, is not confirmed in our calculations.

The stabilization of atomic intercalated carbon crystals is realized from delicate balance of bonding.
As shown in the valence charge density, the formation of strongly polarized $\sigma$ type bonds between Na or Mg and C$_{A}$
and the strength enhanced $\pi$ bonds seems to stabilize the structure of NaC$_{2}$ and MgC$_{2}$.
The discriminative stability seems to be realized by the properous atomic radii of X
and the completion or near completion of the electronic shell of the each C atom in the $K_{4}$ type flame by X=IA or IIA.
Similar mechanism of the stability is expected to exist in the Si and XSi$_{2}$ systems.

Figure 6 shows X-ray diffraction (XRD) patterns of the fully optimized crystal structure of C, NaC$_{2}$, MgC$_{2}$, Si and SrSi$_{2}$.
Monochromatic radiation with wave length 1.541 \AA\ is assumed in these calculations.
As shown in this figure, the compounds have specific XRD peaks at larger $2\theta$ compared with the pure $K_{4}$ crystals.

Figure 7 shows the electron band structures, DOS, and local DOS of valence electrons
for the atoms composing $K_{4}$ type flame and the intercalated atoms
in the fully optimized C, NaC$_{2}$ and MgC$_{2}$ and the Si and SrSi$_{2}$ crystals, respectively.
From the bottom energy levels to the upper levels,
the angular momentum for the electronic states successively changes 
in the order of \textit{s}, \textit{p} and \textit{d} character,
although every component appears in most states.
From the bottom energy levels to those around the Fermi level, the band structures are similar in these crystals.
NaC$_{2}$ and MgC$_{2}$ $K_{4}$ show a metallic and semi conducting feature, respectively, while carbon $K_{4}$ is metallic.
As shown in Fig. 6, both Si and SrSi$_{2}$ have metallic features.
It is noteworthy that the previous first principles calculations for SrSi$_{2}$\cite{SrSi2-AIST-NIMS}
also show qualitatively the same result with the present calculation.

\section{Conclusions and Summary}
The stability of the atomic intercalated carbon $K_{4}$ crystals, XC$_{2}$
(X=H, Li, Be, B, C, N, O, F, Na, Mg, Al, Si, P, S, Cl, K, Ca, Ga, Ge, As, Se, Br, Rb, and Sr),
are evaluated from the geometry optimization and the frozen phonon analysis based on the first principles calculations.
NaC$_{2}$ and MgC$_{2}$ are found to be stable although C $K_{4}$ is not stable.
The formation of strongly polarized $\sigma$ type bonds between Na or Mg and C$_{A}$
and the strength enhanced $\pi$ bonds seems to stabilize the structure of NaC$_{2}$ and MgC$_{2}$.
The discriminative stability seems to be realized by the properous atomic radii of X
and the completion or near completion of the electronic shell of the each C atom in the $K_{4}$ type flame by X=IA or IIA.
The NaC$_{2}$ and MgC$_{2}$ are found to be metallic and semi conducting, respectively.

\section{Appendix}
The values presented in Figure 2 are listed in Table I and II.

\begin{center}
{\bf Acknowledgements}
\end{center}
The authors acknowledge staff of the CCMS, IMR for allowing the use of the HITACHI SR11000 supercomputing facilities.
The authors acknowledge Professor Motoko Kotani and Professor Toshikazu Sunada for information about the crystal structure of SrSi$_{2}$.
We also acknowledge financial support from the CREST, MEXT, and WPI.

\begin{figure*}
\begin{center}
\includegraphics[width=12cm]{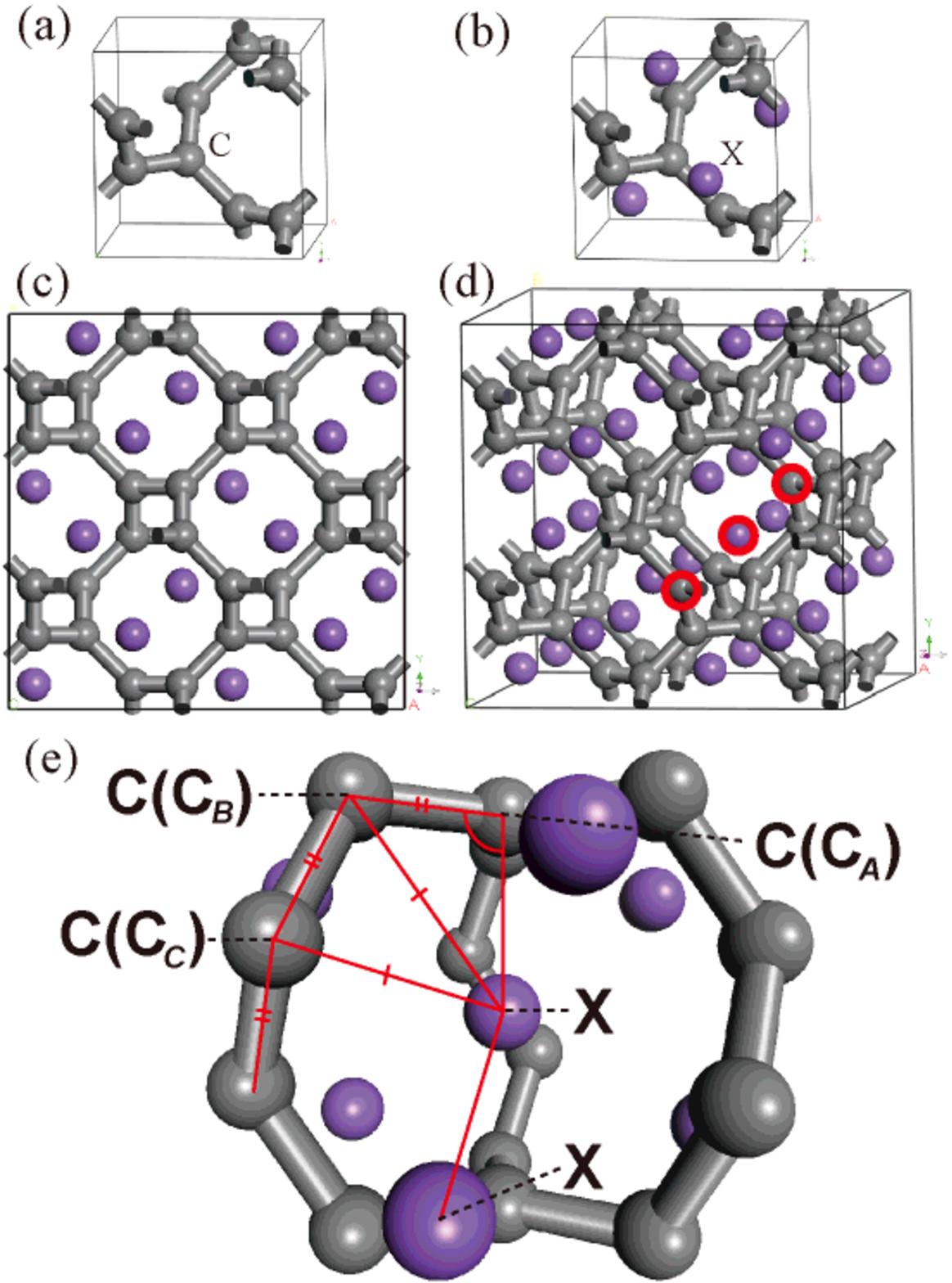}
\caption{\label{fig:K4-XC2-initial-structures-with-extraction}
(a) The conventional unit cell of the carbon $K_{4}$ crystal with $I4_{1}32(O^{8})$ symmetry.
(b) Initial structure of the primitive unit cell of XC$_{2}$ crystal with $K_{4}$ flame of C with $P4_{3}32(O^{6})$ symmetry.
(c) 2$\times$2$\times$2 super cell of the (b).
The symmetric plane of the impurity intercalated system.
(d) Three adjacent atoms (C-X-C) that aligne in a straight line with equivalent distance (C-X and X-C) are encircled with red thick line in (c).
(e) An extracted structure. The C atoms can be classified into C$_A$, C$_B$ and C$_C$.
}
\end{center}
\end{figure*}

\begin{figure*}
\begin{center}
\includegraphics[width=12cm]{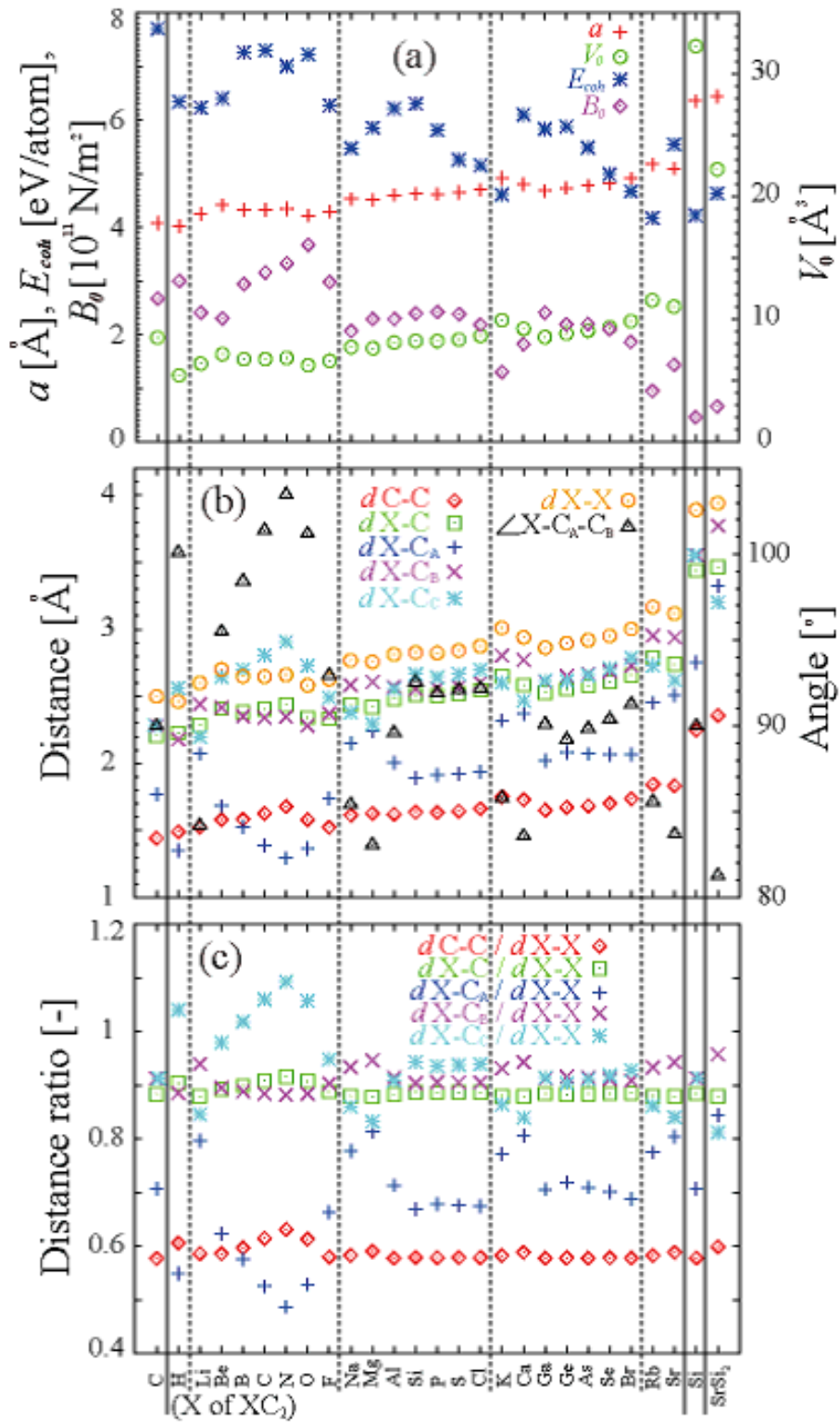}
\caption{\label{fig:K4-C-XC2-Si-SrSi2-structural-detail}
(a) Determined lattice constant($a$) , volume at $a$($V_{0}$),
cohesive energy($E_{coh}$), and bulk modulus at $V_{0}$($B_{0}$), and
(b) the nearest-neighbour distances and angles
for the fully optimized C, XC$_2$, Si and SrSi$_2$ crystal structures with $K_4$ type flame within LDA.
The classification of the C(Si) atoms (C$_A$, C$_B$, and C$_C$)) is based on the irreducibility of the crystal symmetry as shown in Fig. 1.
(c) The distance ratios.
}
\end{center}
\end{figure*}

\begin{figure*}
\begin{center}
\includegraphics[width=12cm]{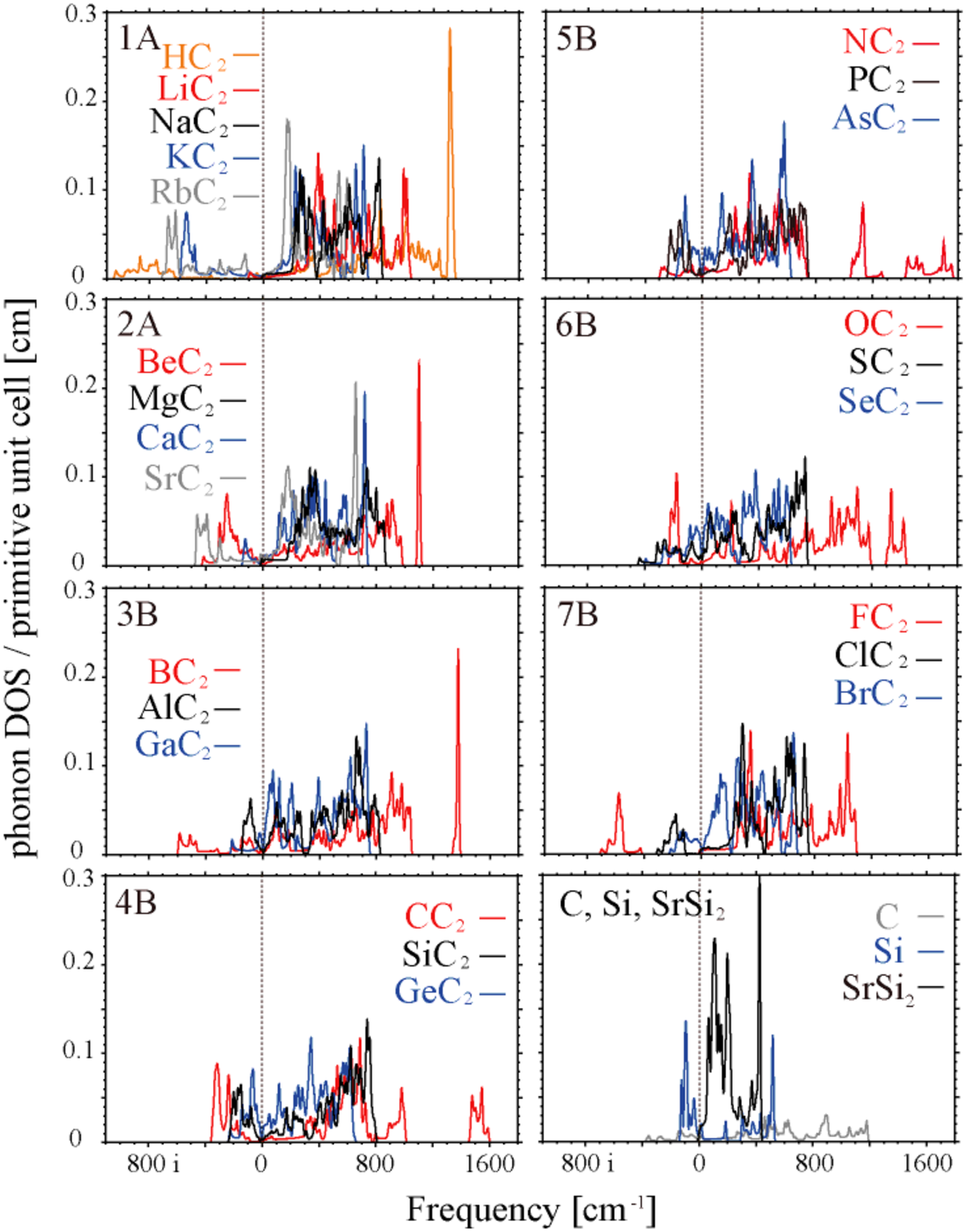}
\caption{\label{fig:K4-XC2-C-Si-SrSi2-sym-phonon-dos}
Phonon density of states(DOS) of XC$_{2}$ (X=H, Li, Be, B, C, N, O, F, Na, Mg, Al, Si, P, S, Cl, K, Ca, Ga, Ge, As, Se, Br, Rb and Sr),
C, Si and SrSi$_{2}$ crystal with $K_{4}$ type flame of C, or Si.
The structures are obtained from geometry optimization under symmetry constraint to the initial structures.
}
\end{center}
\end{figure*}

\begin{figure*}
\begin{center}
\includegraphics[width=17cm]{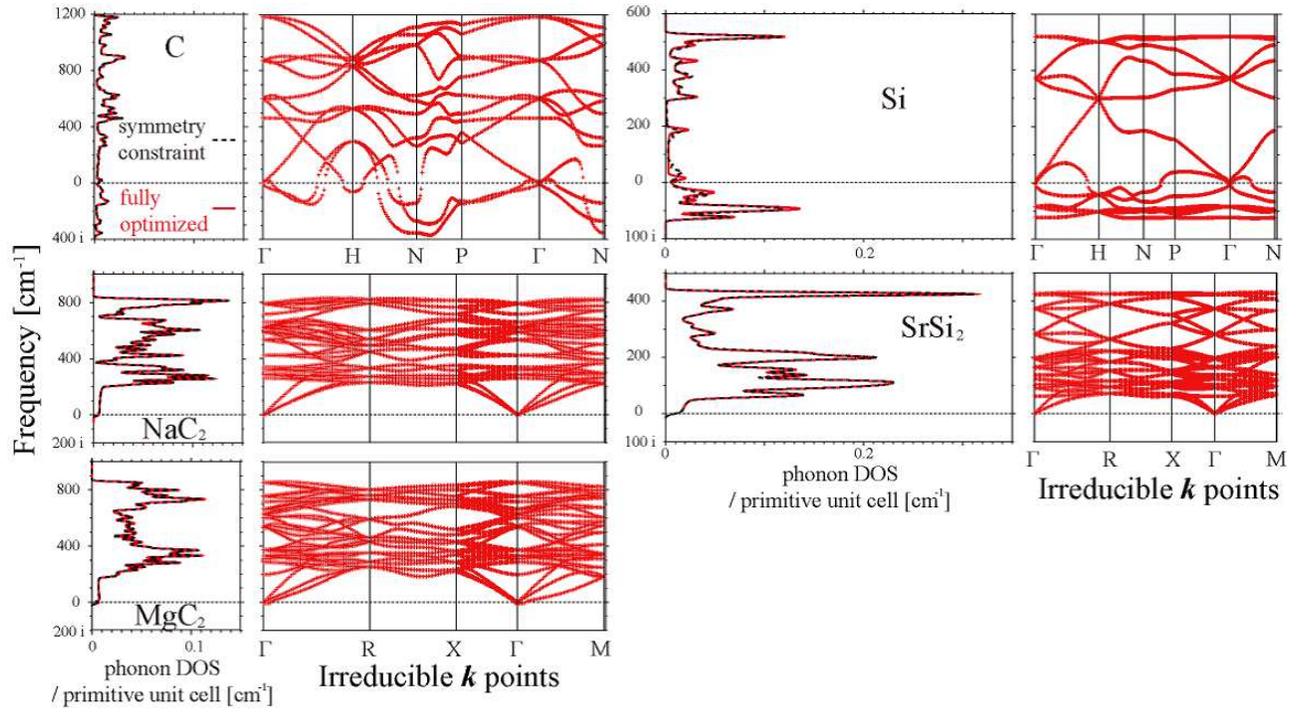}
\caption{\label{fig:K4-C-NaC2-MgC2-Si-SrSi2-sym-non-sym-phonon-dos-band}
(Left) Phonon density of states (DOS) of partially and fully optimized structures
of C, NaC$_{2}$, MgC$_{2}$, Si and SrSi$_{2}$ with $K_{4}$ type flame of C or Si.
(Right) The phonon dispersion relationship for the fully optimized structures.
}
\end{center}
\end{figure*}

\begin{figure*}
\begin{center}
\includegraphics[width=17cm]{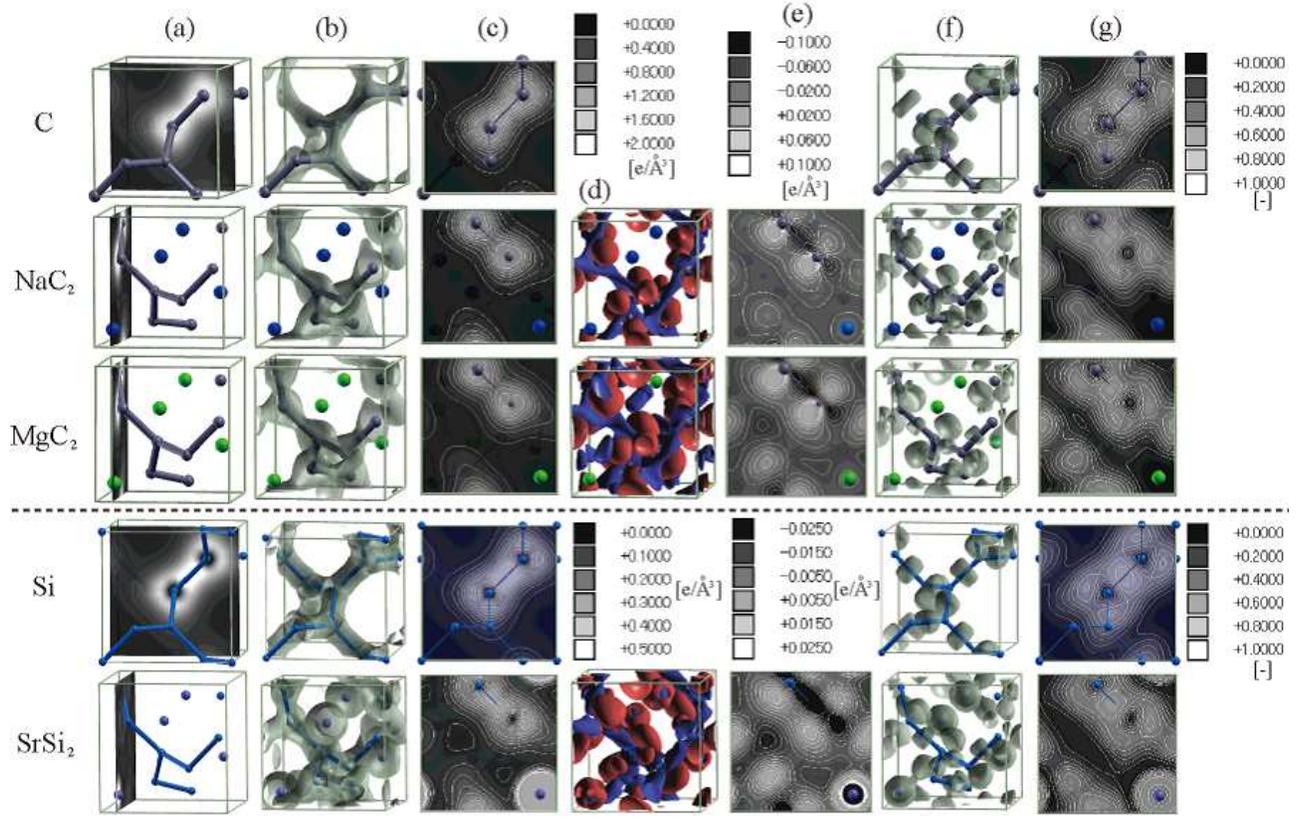}
\caption{\label{fig:K4-C-NaC2-MgC2-Si-SrSi2-chg-diffchg-elf}
(a) The conventional or the primitive unit cells for the fully optimized C, NaC$_2$, MgC$_2$, Si and SrSi$_2$ crystals
with $K_4$ type of flame of C or Si.
The planes are selected to show the distribution of the valence charge density,
difference between the valence charge density of the crystals (ex. $\rho_{NaC_{2}}(\textbf{r})$) and their separated components:
the intercalated atoms (ex. $\rho_{Na}(\textbf{r})$) and the flame (ex. $\rho_{C_{2}}(\textbf{r})$),
and the electronic localization function (ELF).
The isosurfaces of the valence charge density (1.5, 1.0, 1.0, 0.35 and 0.35 $e$/\AA $^3$
for C, NaC$_2$, MgC$_2$, Si and SrSi$_2$, respectively) (b) and the contours (c).
The isosurfaces of the differences of the valence charge density
(0.05, 0.05 and 0.025 +(-)$e$/\AA $^3$ coloured with red(blue) for NaC$_2$, MgC$_2$ and SrSi$_2$, respectively) (d)
and the contours (e).
The isosurfaces of the ELF (0.8 [-]) (f) and the contours (g).
For (c) and (e), the high population around the Sr atoms is neglected for easier comparison.
{XCrySDen}\cite{Kokalj} was used for the visualization.
}
\end{center}
\end{figure*}

\begin{figure*}
\begin{center}
\includegraphics[width=12cm]{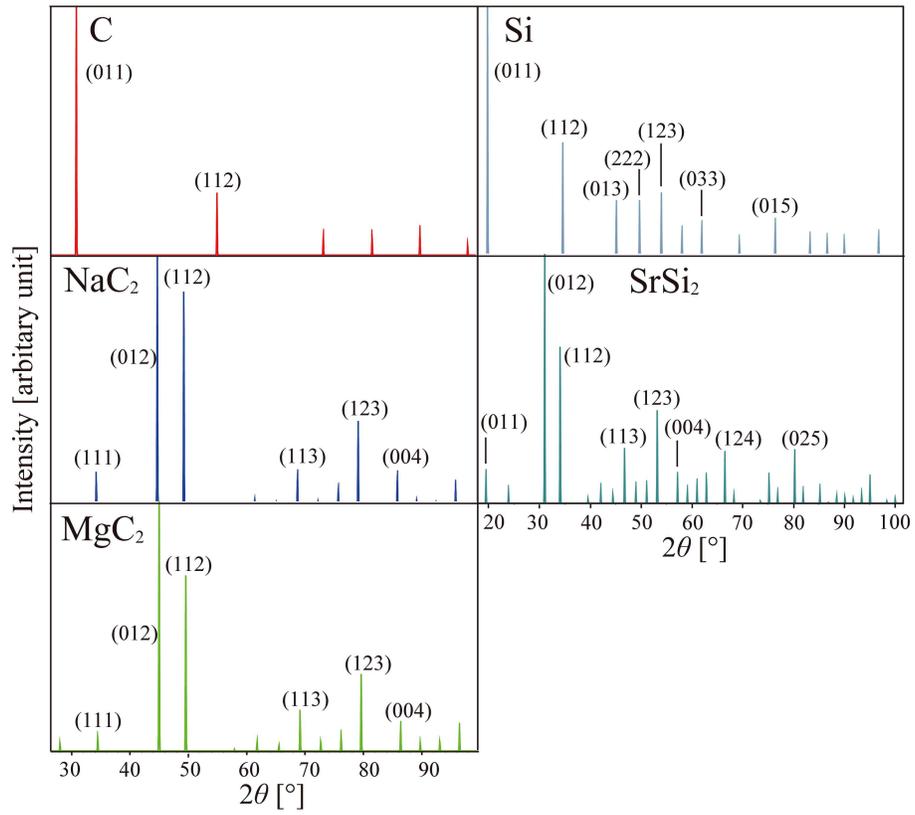}
\caption{\label{fig:K4-C-NaC2-MgC2-Si-SrSi2-XRD}
XRD patterns for the fully optimized structures of C, NaC$_{2}$, MgC$_{2}$, Si and SrSi$_{2}$
with $K_{4}$ type flame of C or Si.
}
\end{center}
\end{figure*}

\begin{figure*}
\begin{center}
\includegraphics[width=12cm]{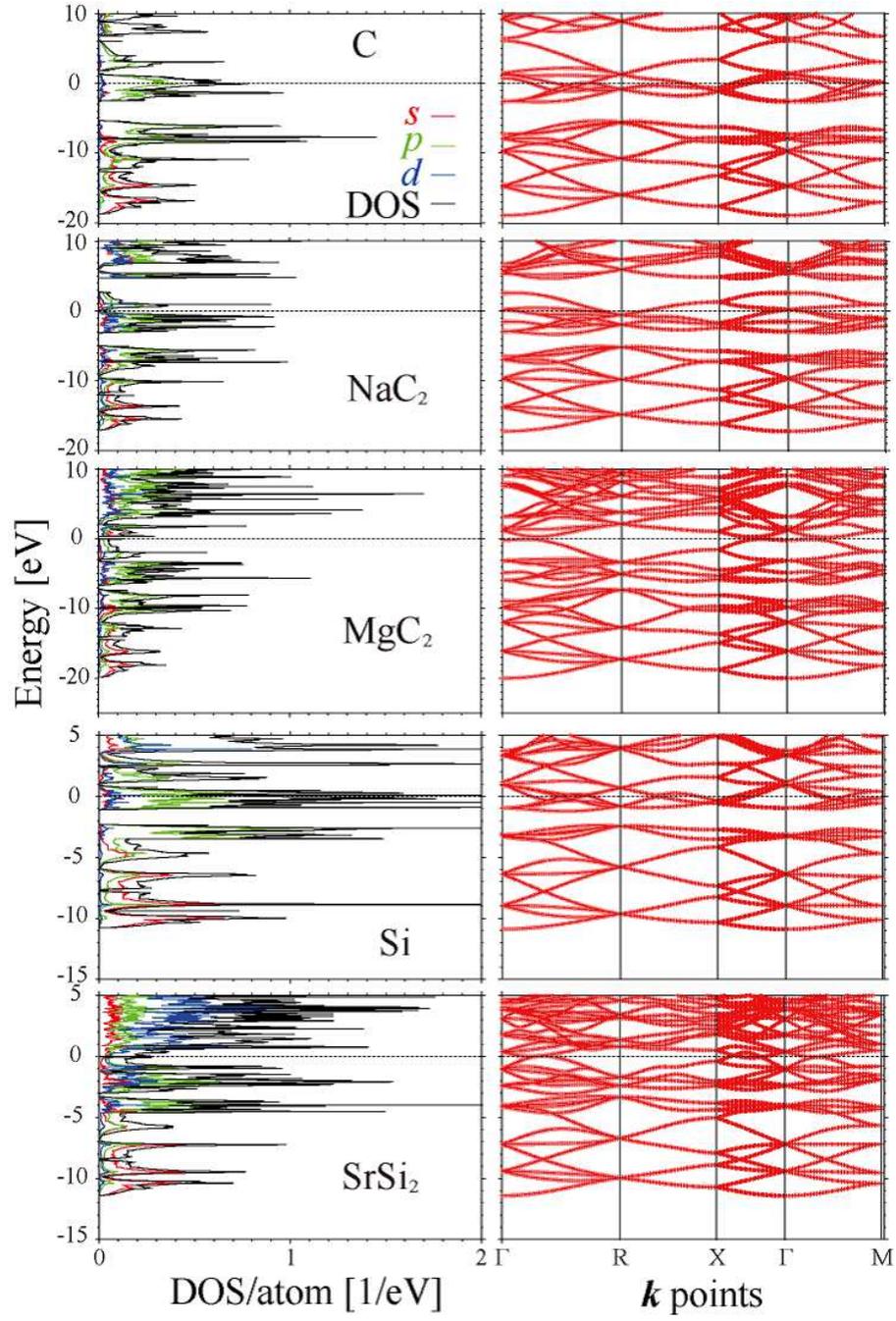}
\caption{\label{fig:K4-C-NaC2-MgC2-Si-SrSi2-KS-dos-band}
Electronic density of states (DOS) and the dispersion relationship for the fully optimized conventional or primitive unit cells
of C, NaC$_2$, MgC$_2$, Si and SrSi$_2$ crystals with $K_4$ type flame of C or Si.
}
\end{center}
\end{figure*}

\begin{table*}
\begin{center}
\caption{\label{tab:calc}
Determined lattice constant($a$), volume at $a$($V_{0}$),
cohesive energy($E_{coh}$), and bulk modulus at $V_{0}$($B_{0}$)
for the fully optimized C, XC$_2$, Si and SrSi$_2$ crystal structures with $K_4$ type flame within LDA.
}
\begin{ruledtabular}
\begin{tabular}{cccccc}
Component&$a$ [\AA\ ]&$V_0$ [\AA\ $^{3}$/atom]&$E_{coh}$ [eV/atom]& $B_0$ [10$^{11}$ $\frac{N}{m^2}$]\\
\hline
\hline
C\cite{IKNSKA-PRL-2009}&4.082 &8.502 &7.702 &2.67 \\
\hline
\hline
HC$_{2}$&4.025 &5.433 &6.335 &3.00 \\
\hline
\hline
LiC$_{2}$&4.250 &6.396 &6.231 &2.41 \\
\hline
BeC$_{2}$&4.412 &7.158 &6.401 &2.30 \\
\hline
BC$_{2}$&4.328 &6.758 &7.259 &2.94 \\
\hline
CC$_{2}$&4.326 &6.747 &7.295 &3.16 \\
\hline
NC$_{2}$&4.348 &6.850 &7.003 &3.33 \\
\hline
OC$_{2}$&4.218 &6.252 &7.221 &3.67 \\
\hline
FC$_{2}$&4.296 &6.607 &6.273 &2.98 \\
\hline
\hline
NaC$_{2}$&4.526 &7.728 &5.476 &2.06 \\
\hline
MgC$_{2}$&4.506 &7.623 &5.856 &2.29 \\
\hline
AlC$_{2}$&4.598 &8.102 &6.214 &2.29 \\
\hline
SiC$_{2}$&4.622 &8.227 &6.301 &2.39 \\
\hline
PC$_{2}$&4.619 &8.214 &5.805 &2.42 \\
\hline
SC$_{2}$&4.644 &8.345 &5.261 &2.38 \\
\hline
ClC$_{2}$&4.699 &8.646 &5.152 &2.18 \\
\hline
\hline
KC$_{2}$&4.920 &9.926 &4.606 &1.30 \\
\hline
CaC$_{2}$&4.803 &9.232 &6.099 &1.82 \\
\hline
GaC$_{2}$&4.679 &8.536 &5.831 &2.41 \\
\hline
GeC$_{2}$&4.731 &8.823 &5.875 &2.19 \\
\hline
AsC$_{2}$&4.773 &9.059 &5.485 &2.20 \\
\hline
SeC$_{2}$&4.823 &9.351 &4.987 &2.10 \\
\hline
BrC$_{2}$&4.908 &9.852 &4.668 &1.86 \\
\hline
\hline
RbC$_{2}$&5.175 &11.550 &4.172 &0.95 \\
\hline
SrC$_{2}$&5.096 &11.030 &5.545 &1.44 \\
\hline
\hline
Si&6.354 &32.089 &4.220 &0.46 \\
\hline
\hline
SrSi$_{2}$&6.436 &22.216 &4.634 &0.66 \\
\end{tabular}
\end{ruledtabular}
\end{center}
\end{table*}

\begin{table*}
\begin{center}
\caption{\label{tab:d-angle-detail}
Determined nearest-neighbour distances and angles
for the fully optimized C, XC$_2$, Si and SrSi$_2$ crystal structures with $K_4$ type flame within LDA.
The classification of the C(Si) atoms (C$_A$, C$_B$ and C$_C$)) is based on the irreducibility of the crystal symmetry as shown in Fig. 1.
The smallest dihedral angles for the nearest-neighbor C(Si) atoms of $K_{4}$ flame are also shown.
}
\begin{ruledtabular}
\begin{tabular}{ccccccccccc}
Component&	$d$ C-C [\AA\ ]&$\overline{d}$ X-C [\AA\ ]&$d$ X-C$_A$ [\AA\ ]&	$d$ X-C$_B$ [\AA\ ]&	$d$ X-C$_C$ [\AA\ ]& $d$ X-X [\AA\ ]& $\angle$ X-C$_A$-C$_B$ [$^{\circ}$]& $\angle$ C-C-C-C [$^{\circ}$]\\
\hline
\hline
C\cite{IKNSKA-PRL-2009}&	1.443& 2.209&	1.768&	2.282&	2.282&	2.500&	90.000& 70.529\\
\hline
\hline
HC$_2$&	1.492&	2.226& 1.350&	2.180&	2.563&	2.463&	100.081& 67.946\\
\hline
\hline
LiC$_2$&	1.525&	2.286& 2.071&	2.444&	2.199&	2.601&	84.183& 69.686\\
\hline
BeC$_2$&	1.581&	2.409& 1.683&	2.417&	2.644&	2.701&	95.477& 69.782\\
\hline
BC$_2$&	1.582&	2.383& 1.526&	2.354&	2.698&	2.650&	98.434& 68.732\\
\hline
CC$_2$&	1.628&	2.405& 1.391&	2.341&	2.808&	2.649&	101.393& 67.330\\
\hline
NC$_2$&	1.679&	2.437& 1.295&	2.347&	2.908&	2.661&	103.471& 67.202\\
\hline
OC$_2$&	1.582&	2.343& 1.365&	2.282&	2.729&	2.582&	101.180& 65.807\\
\hline
FC$_2$&	1.524&	2.333& 1.741&	2.372&	2.492&	2.628&	92.955& 70.313\\
\hline
\hline
NaC$_2$&	1.614&	2.436& 2.152&	2.585&	2.381&	2.769&	85.408& 70.005\\
\hline
MgC$_2$&	1.628&	2.422& 2.244&	2.608&	2.295&	2.757&	83.057& 69.323\\
\hline
AlC$_2$&	1.624&	2.483& 2.006&	2.572&	2.554&	2.813&	89.592& 70.525\\
\hline
SiC$_2$&	1.637& 2.508& 1.890&	2.555&	2.666&	2.827&	92.542& 70.369\\
\hline
PC$_2$&	1.634&	2.503& 1.916&	2.559&	2.643&	2.825&	91.904& 70.439\\
\hline
SC$_2$&	1.645&	2.520& 1.920&	2.573&	2.666&	2.843&	92.100& 70.420\\
\hline
ClC$_2$&	1.663&	2.548& 1.938&	2.601&	2.698&	2.874&	92.170& 70.412\\
\hline
\hline
KC$_2$& 1.752&	2.649& 2.322&	2.804&	2.602&	3.011&	85.787& 70.088\\
\hline
CaC$_2$&	1.730& 2.584& 2.369&	2.773&	2.466&	2.940&	83.574& 69.498\\
\hline
GaC$_2$&	1.653&	2.530& 2.020&	2.613&	2.617&	2.863&	90.101& 70.529\\
\hline
GeC$_2$&	1.673&	2.556& 2.082&	2.653&	2.617&	2.896&	89.211& 70.513\\
\hline
AsC$_2$&	1.685&	2.578& 2.071&	2.667&	2.659&	2.919&	89.841& 70.528\\
\hline
SeC$_2$&	1.704&	2.608& 2.069&	2.689&	2.707&	2.951&	90.387& 70.525\\
\hline
BrC$_2$&	1.736&	2.659& 2.067&	2.728&	2.787&	3.004&	91.257& 70.490\\
\hline
\hline
RbC$_2$&	1.845&	2.786& 2.453&	2.954&	2.729&	3.167&	85.575& 70.042\\
\hline
SrC$_2$&	1.834&	2.740& 2.507&	2.939&	2.619&	3.118&	83.684& 69.533\\
\hline
\hline
Component&	$d$ Si-Si [\AA\ ]&	$\overline{d}$ X-Si [\AA\ ] & $d$ X-Si$_A$ [\AA\ ]&	$d$ X-Si$_B$ [\AA\ ]&	$d$ X-Si$_C$ [\AA\ ]& $d$ X-X [\AA\ ]&	$\angle$ X-Si$_A$-Si$_B$ [$^{\circ}$]& $\angle$ Si-Si-Si-Si [$^{\circ}$]\\
\hline
\hline
Si&	2.247&	3.439& 2.752&	3.553&	3.553&	3.891&	90.000& 70.529\\
\hline
\hline
SrSi$_2$&	3.463& 2.358&	3.324&	3.772&	3.200&	3.941&	81.270& 68.605\\
\end{tabular}
\end{ruledtabular}
\end{center}
\end{table*}


\newpage

\begin{thebibliography}{99}
\bibitem{Sunada}
T. Sunada, Not. Amer. Math. Soc., \textbf{55}, 208 (2008).
\bibitem{Kawazoe}
Y. Kawazoe, T. Kondow, and K. Ohno, Eds., \textit{Clusters and Nanomaterials}., Springer (2002), and references therein.
\bibitem{RC-PRB-2008}
G.-M. Rignanese and J.-C. Charlier, Phys. Rev. B., \textbf{78}, 125415 (2008).
\bibitem{IKNSKA-PRL-2009}
M. Itoh, M. Kotani, H. Naito, T. Sunada, Y. Kawazoe, and T. Adschiri, Phys. Rev. Lett., \textbf{102}, 055703 (2009).
\bibitem{Yao}
Y. Yao, J.S. Tse, J. Sun, D.D. Klug, R. Marto\v{n}\'{a}k, and T. Iitaka, Phys. Rev. Lett., \textbf{102}, 229601 (2009).
\bibitem{SrSi2}
K. Janzon, H. Sch\"{a}fer, and A. Weiss, Angew. Chem. Int. Ed., \textbf{4}, 245 (1965).
\bibitem{carbides-review}
U. Ruschewitz, Coord. Chem. Rev., \textbf{244}, 115 (2003). 
\bibitem{HK}
P. Hohenberg, and W. Kohn, Phys. Rev., \textbf{136}, B 864 (1964).
\bibitem{KS}
W. Kohn, and L. J. Sham, Phys. Rev., \textbf{140}, A 1133 (1965).
\bibitem{Kresse}
G. Kresse, and J. Furthm\"uller, Phys. Rev. B., \textbf{54}, 11169 (1996).
\bibitem{PZ}
J. P. Perdew, and A. Zunger, Phys. Rev. B., \textbf{23}, 5048 (1981).
\bibitem{CA}
D. M. Ceperley, and B. J. Alder, Phys. Rev. Lett., \textbf{45}, 566 (1980).
\bibitem{PAW}
P. E. Bl\"{o}chl, Phys. Rev. B., \textbf{50}, 17953 (1994).
\bibitem{Murnaghan}
F. D. Murnaghan, Proc. Natl. Acad. Sci. USA., \textbf{30}, 244 (1944).
\bibitem{FROPHO}
A. Togo, FROPHO, http://fropho.sourceforge.net/
\bibitem{Parlinski-Li-Kawazoe}
K. Parlinski, Z. Q. Li, and Y. Kawazoe, Phys. Rev. Lett., \textbf{78}, 4063 (1997).
\bibitem{SrSi2-AIST-NIMS}
Y. Imai, and A. Watanabe, Intermetallics, \textbf{10}, 333 (2002).
\bibitem{Tohei}
T. Tohei, A. Kuwabara, F. Oba, and I. Tanaka, Phys. Rev. B., \textbf{73}, 064304 (2006).
\bibitem{Giannozzi}
P. Giannozzi, S. Gironcoli, P. Pavone, and S. Baroni, Phys. Rev. B., \textbf{43}, 7231 (1991).
\bibitem{Cohen}
M. L. Cohen, Science, \textbf{234}, 4776 (1986).
\bibitem{Kokalj}
A. Kokalj, Comp. Mat. Sci., \textbf{28}, 155 (2003).
\end{thebibliography}
\end{document}